\documentclass[nofootinbib,amsmath,preprint,floatfix,showpacs,prl]{revtex4}
\usepackage{graphics}
\usepackage{longtable}
\usepackage{bm}
\usepackage{epsfig}

\newcommand{\ket}[1]{\mbox{$| #1 \rangle$}}

\newcommand{\scaldir}[2]{\mbox{$\langle #1 | #2 \rangle$}}
\newcommand{\mteldir}[3]{\mbox{$\langle #1 | #2 | #3 \rangle$}}

\begin{document}

\title{Environment assisted electron capture}
\author{Kirill Gokhberg and Lorenz S. Cederbaum}
\affiliation{Theoretische Chemie, Physikalisch-Chemisches Institut,
Universit\"at Heidelberg, Im Neuenheimer Feld 229, D-69120 Heidelberg, Germany}

\begin{abstract}
Electron capture by {\it isolated} atoms and ions proceeds by photorecombination. 
In this process a species captures a free electron by emitting a photon 
which carries away the excess energy. It is shown here that in the 
presence of an {\it environment} a competing non-radiative electron capture 
process can take place due to long range electron correlation. In this interatomic (intermolecular) process the 
excess energy is transferred to neighboring species. The asymptotic 
expression for the cross section of this process is derived. We 
demonstrate by explicit examples that under realizable conditions
the cross section of this 
interatomic process can clearly dominate that of photorecombination.
\end{abstract}

\maketitle

Electron capture by atoms and ions is a basic phenomenon of general interest. 
If the atom or ion is isolated in space, the only possibility to capture a 
free electron is by photorecombination (PR) \cite{bk:sobelman1972}. This process 
can be viewed as the inverse of photoionization or photodetachment. In a collision 
event, a free electron of energy $\varepsilon$ is captured into a bound level of 
a species A which is predominantly the ground state \cite{mueller2008}, while a photon 
is simultaneously emitted which carries away the excess 
energy. For neutral species A this process is commonly called radiative attachment, 
while if A is a positive ion the term radiative recombination is used. 
Photorecombination has been widely studied in media with low atomic densities 
\cite{mueller2008,herbst1981,flannery2006,bk:osterbrock2006,ichihara2000}
where it is the dominant electron capture mechanism.

In this work we enrich the physics of electron capture by discussing what happens 
when there are some other species in the neighborhood of A. 
We show that due to the presence of these neighbors a new non-radiative interatomic electron capture 
mechanism becomes operative. This mechanism is illuminated and found to emerge from 
long-range electron correlation. We demonstrate by explicit examples that the environment 
can strongly enhance the electron capture by atoms and ions compared to the capture by the 
isolated species. The dependence of this enhancement on the properties of the A and of its 
neighbors as well as on the interatomic distances between them is discussed.

\begin{figure}[htb]
\begin{center}
\includegraphics[width=12.cm]{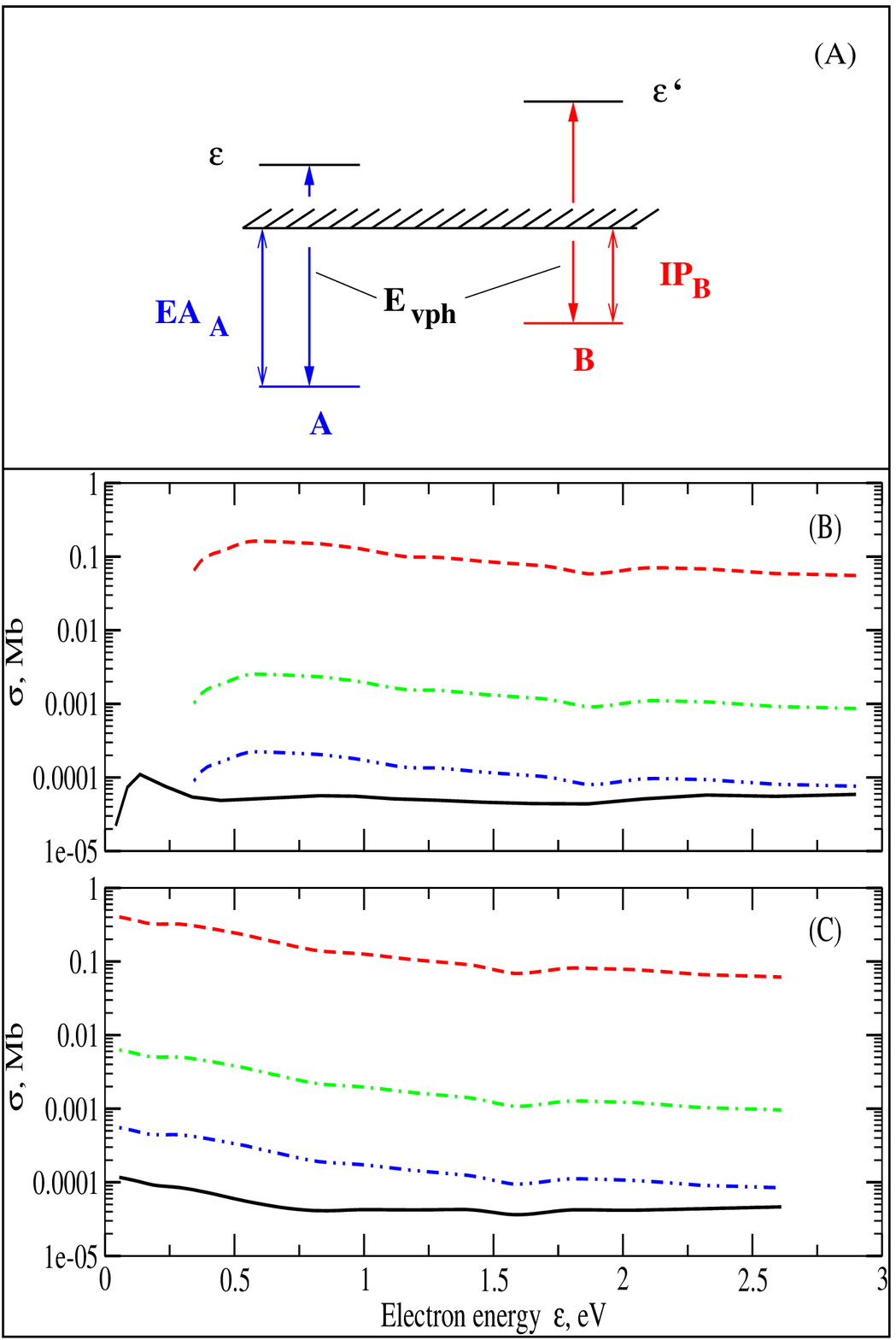}
\caption{(A) Schematic representation of the ICEC process (see the text for explanations).
(B) PR cross section of Br atom and ICEC cross sections of Br with Cl$^-$ as its neighbor at different interatomic
separations: solid line - $\sigma_{\rm PR}^{\rm Br}(\varepsilon)$; dashed line - $\sigma_{\rm ICEC}$ at {\it R}=1 nm;
dashed-dotted line-$\sigma_{\rm ICEC}$ at {\it R}=2 nm; dashed-double dotted line - $\sigma_{\rm ICEC}$ at {\it R}=3 nm. 
(C) PR cross section of Cl atom and ICEC cross sections of Cl with Br$^-$ as its neighbor at different interatomic
separations: solid line - $\sigma_{\rm PR}^{\rm Cl}(\varepsilon)$; dashed line - $\sigma_{\rm ICEC}$ at {\it R}=1 nm;
dashed-dotted line - $\sigma_{\rm ICEC}$ at {\it R}=2 nm; dashed-double dotted line - $\sigma_{\rm ICEC}$ at {\it R}=3 nm.}
\label{fg1}
\end{center}
\end{figure}

Let us begin with an isolated atom A. In the PR process A captures a 
free electron of energy $\varepsilon$ and the excess energy is released by emitting 
a photon of energy $h\nu$ = EA$_{\rm A}$+$\varepsilon$, where EA$_{\rm A}$ is the binding energy of 
the excess electron (electron affinity of A if A is a neutral atom and 
ionization potential of A if A is an ion). For convenience we shall 
address EA$_{\rm A}$ in the following as the electron affinity of A. Next, we 
consider an atom A in the presence of a neighboring atom B. Now A can 
capture a free electron of energy $\varepsilon$ and utilize the excess energy to 
ionize its neighbor B. Energy conservation tells us that EA$_{\rm A}$+$\varepsilon$=IP$_{\rm B}$+$\varepsilon^\prime$,
where IP$_{\rm B}$  stands for the energy needed to remove an 
electron from B, while $\varepsilon^\prime$ designates the energy of the outgoing 
electron. For simplicity we call IP$_{\rm B}$ the ionization potential of B even 
if B is not a neutral atom. The electron is captured by A, while in a 
correlated step B is ionized, and the excess energy is carried away not 
by a photon but by the outgoing electron. This process occurs due to 
electron correlation in the system induced by the Coulomb interaction, 
and we call it interatomic Coulombic electron capture (ICEC). The 
process is schematically shown in Fig.\ref{fg1}A. ICEC is an elementary process 
totally different from PR, and since they might be present 
simultaneously and compete, we are interested in comparing their 
characteristics. The threshold for PR lies at $\varepsilon$=0, while the threshold 
$\varepsilon^t$ for ICEC depends on the neighbor B. In systems where IP$_{\rm B}$$<$ EA$_{\rm A}$ 
this threshold is also at 0. If IP$_{\rm B}$$>$ EA$_{\rm A}$, the ICEC threshold energy 
is $\varepsilon^t$=IP$_{\rm B}$-EA$_{\rm A}$ and slower electrons will be captured exclusively by 
the PR mechanism, while at  $\varepsilon>\varepsilon^t$ both processes are operative. Which 
process, PR or ICEC, will dominate? To answer this question we must 
compare the cross sections of these processes. While that of PR is 
generally known, we discuss below the cross section of ICEC.

The ICEC process described above can be treated by the methods of 
multichannel scattering \cite{bk:taylor2006}. We first construct the initial and final 
states of the system using one-electron orbitals and energies obtained, 
e.g. by solving Hartree-Fock equations. The bound orbitals and energies 
are denoted \ket{\gamma l m \mu} and $\varepsilon_{\gamma l m}$, where {\it l}, {\it m} $\mu$, and $\gamma$ stand for 
the orbital angular momentum, its projection on the interatomic axis, projection of the electron spin, and the rest 
of quantum numbers, respectively. These 
orbitals satisfy the usual normalization conditions 
$\scaldir{\gamma^{\prime}l^{\prime}m^{\prime}\mu^{\prime}}{\gamma lm\mu}
=\delta_{\gamma\gamma^{\prime}}\delta_{ll^{\prime}}\delta_{mm^{\prime}}\delta_{\mu\mu^{\prime}}$. 
We designate continuum one-electron states as \ket{{\bf k} \mu}  
with corresponding energies $\varepsilon_{\bf k}$, where {\bf k} is the wavevector of the 
continuum electron, and assume the normalization condition $\scaldir{{\bf k}^\prime \mu^\prime}{{\bf k} \mu}=
\delta({\bf k}^\prime-{\bf k})\delta_{\mu\mu^{\prime}}$. Using these orbitals one can construct the one-determinant 
ground state \ket{\Phi_{0}} of A+B. For an incoming electron of momentum {\bf k} the 
initial state is given by $\ket{\Phi_{\bf k}}=\hat{c}^{\dagger}_{\bf k}\ket{\Phi_{0}}$, where the operator 
$\hat{c}^{\dagger}_{\bf k}$
creates an electron in the continuum orbital \ket{{\bf k} \mu}. In the out channel 
the captured electron is in the lowest unoccupied orbital of A denoted 
by \ket{a_{A}l_{A}m_{A}\mu_{A}} and an electron is emitted from the orbital \ket{j_{B}l_{B}m_{B}\mu_{B}} of 
B into the continuum orbital \ket{{\bf k^\prime} \mu^\prime}. 
The corresponding final state is
$\ket{\Phi_{{\bf k^\prime}a_{A}j_{B}}}=\hat{c}^{\dagger}_{\bf k^\prime}\hat{c}^{\dagger}_{a_{A}}
\hat{c}_{j_{B}}\ket{\Phi_{0}}$, where ${c}^{\dagger}_{a_{A}}$ is a creation operator adding an electron to 
\ket{a_{A}l_{A}m_{A}\mu_{A}} and $\hat{c}_{j_{B}}$ is a destruction operator removing an electron from 
\ket{j_{B}l_{B}m_{B}\mu_{B}}.

The scattering potential coupling between the initial and final states is the 
electron-electron interaction $\hat{V}=1/2\sum_{i\neq j}e^2/|{\bf r_{i}}-{\bf r_{j}}|$, where the sum 
runs over the coordinates of all electrons in the e+A+B system. The 
on-shell scattering matrix reads $t(out\leftarrow in)=\mteldir{\Phi_{{\bf k^\prime}a_{A}j_{B}}}{\hat{V}}{\Phi_{\bf k}+}$, 
where \ket{\Phi_{\bf k}+} is the outgoing scattering state \cite{bk:taylor2006}. Knowing the
on-shell t-matrix one finds the differential cross section
\begin{equation}
\frac{d\sigma({\bf k^\prime}\leftarrow {\bf k})}{d\Omega_{\bf k^\prime}}=\frac{m^2_e}{(2\pi)^2\hbar^4}\frac{k^\prime}{k}
|t(out\leftarrow in)|^2,
\label{eq1}
\end{equation}
where the absolute values of the wavevectors of the initial and final 
states are connected by energy conservation. To obtain the total cross 
section for ICEC, the differential cross section is summed over all 
final states and averaged over all initial states at the energy $\varepsilon= k^2/2m_e$ giving
\begin{equation}
\sigma_{\rm ICEC}(k)=\frac{1}{g_{in}}
\sum_{in,out}\int \frac{d\Omega_{\bf k}}{4\pi}\int d\Omega_{\bf k^\prime}\frac{d\sigma({\bf k^\prime}
\leftarrow {\bf k})}{d\Omega_{\bf k^\prime}},
\label{eq2}
\end{equation}
where $g_{in}$ is the multiplicity of the initial state.

As our aim here is to discuss a process of interatomic nature, we do not 
attempt to calculate $\sigma_{\rm ICEC}$ exactly but concentrate instead on deriving 
an analytical asymptotic expression valid at large distances between A 
and B. This can be done using the framework of the virtual photon transfer 
model, see \cite{averbukh2004} and references therein. Accordingly, as the interatomic distance {\it R} between A and B 
grows, an interatomic process such as ICEC can be viewed as events 
taking place separately on A and B and linked by the transfer of a 
virtual photon. In the first event a free electron is
captured by A and a virtual photon is emitted. In the second event this 
photon is transferred to B and ionizes it. Mathematically this follows 
from the expansion of the scattering potential $\hat{V}$ in inverse powers of 
{\it R}. The leading term of the on-shell t-matrix due to this expansion is 
the interaction between the two transition dipoles corresponding to the 
two events mentioned above and reads
\begin{equation}
t(out\leftarrow in)=\frac{C_S}{R^3}\sum_{m=-1}^{1}B_{m}\mteldir{\Phi_{a_{A}}}{\hat{D}_{m}}{\Phi_{\bf k}+}
\mteldir{\Phi_{0}}{\hat{D}_{m}}{\Phi_{{\bf k^\prime}j_{B}}}^\ast,
\label{eq3}
\end{equation}
where $\ket{\Phi_{a_{A}}}=\hat{c}^{\dagger}_{a_{A}}\ket{\Phi_{0}}$, 
$\ket{\Phi_{{\bf k^\prime}j_{B}}}=\hat{c}^{\dagger}_{\bf k^\prime}\hat{c}_{j_{B}}\ket{\Phi_{0}}$, 
$B_{0}$=-2, $B_{\pm 1}$=1, $C_{S}$ is a coefficient of order unity 
depending on the spin of the initial and final states, and
$\hat{D}_m$ is the m-th component of the dipole transition operator. As {\it R} increases, the matrix elements 
in Eq.(\ref{eq3}) approach the 
matrix elements one would obtain for the isolated species A and B. This 
can be utilized in the asymptotic expression and allows one to
express $\sigma_{\rm ICEC}$ by quantities of the isolated A and B. Inserting the 
resulting t-matrix into Eq.(\ref{eq2}), one can separate the integrations on A 
from those on B and express these integrals by the photorecombination 
cross section $\sigma_{\rm PR}^{\rm (A)}$ of the isolated A species, and the photoionization 
cross section $\sigma_{\rm PI}^{\rm \rm (B)}$ of the isolated B species. The final result takes 
on the following appealing form
\begin{equation}
\sigma_{\rm ICEC}(\varepsilon)=P(E_{vph},R)\sigma_{\rm PR}^{\rm (A)}(\varepsilon),
\label{eq4}
\end{equation}
where $P(E_{vph},R)=3\hbar^4c^4\sigma_{\rm PI}^{\rm (B)}(\varepsilon^{\prime})/2\pi R^6 E_{vph}^4$ is a 
dimensionless coefficient and $E_{vph}$=EA$_{\rm A}$+$\varepsilon$ is the energy of the virtual photon.

This expression provides a direct comparison of the efficiencies of PR 
and ICEC to capture a free electron of energy $\varepsilon$. Once $P(E_{vph},R)$ is 
larger than unity, ICEC is favorable to PR. Three ingredients enter 
this coefficient: the distance {\it R} between A and its neighbor B, the 
energy $E_{vph}$ of the virtual photon, and the ionization cross section of 
B. $P(E_{vph},R)$ grows with decreasing {\it R} as does the interaction between the 
two transition dipoles which is proportional to $R^{-6}$. At a fixed 
interatomic distance the interatomic capture is likely to increase as 
the virtual photon energy $E_{vph}$ decreases. For a given system this 
implies that ICEC is most effective for slow incoming electrons. In 
addition to the fourth power of $E_{vph}$ in the denominator, the energy 
dependence of $P(E_{vph},R)$ is, however, also determined by $\sigma_{\rm PI}^{\rm \rm (B)}(\varepsilon^\prime)$ in 
the numerator. The virtual photon is emitted from atom A and ionizes the 
neighbor B, and thus the ICEC cross section at large separations {\it R} is 
proportional to the photoionization cross section of B. Consequently, 
choosing a neighbor with a high photoionization cross section will 
favor ICEC.

Having discussed the relation between $\sigma_{\rm ICEC}$ and $\sigma_{\rm PR}^{\rm (A)}$ we would like to compare these 
cross sections in some realistic illustrative examples. As the first application we consider a system, where A
is an halogen atom and B is an halide of a different halogen species. 
Let us start with the Br atom and Cl$^{-}$ as its neighbor. Since EA$_{\rm Br}$=3.313 eV and IP$_{{\rm Cl}^{-}}$=3.601 eV
\cite{mandl1976}, the ICEC threshold lies at $\varepsilon$=0.288 eV. Below this energy only PR is possible, whereby 
the electron is captured by Br to form Br$^-$ and a photon of energy
$\varepsilon$+EA$_{\rm Br}$ is emitted. PR cross sections of a species A are usually obtained from the photoionization or 
photodetachment cross sections of A$^{-}$ 
using the principle of detailed balance \cite{bk:sobelman1972}:
$k^2g_{\rm A}\sigma_{\rm PR}^{\rm (A)}(\varepsilon)=k^2_{ph}g_{{\rm A}^-}\sigma_{\rm PI}^{({\rm A}^-)}(\varepsilon)$, 
where {\it k} and {\it k}$_{ph}$ are the absolute values of the wavevectors of the captured electron and emitted photon, 
respectively, while $g_{\rm A}$ and $g_{{\rm A}^-}$ are the statistical weights of the quantum states of A and A$^-$.
To obtain the PR cross section of Br from the detailed balance we used the experimental photodetachment cross 
section of Br$^-$ reported in Ref.\cite{mandl1976}. Since the ground state of Br is of $^2P$ symmetry and the 
ground state of Br$^{-}$ is of $^1S$ symmetry, one has $g_{\rm Br}=6$ and $g_{\rm Br^-}=1$. 
To compute the ICEC cross section we used Eq.(\ref{eq4}) and the experimental photodetachment cross 
section of Cl$^-$ \cite{mandl1976}. Fig.\ref{fg1}B shows $\sigma_{\rm PR}^{\rm (Br)}(\varepsilon)$ and 
$\sigma_{\rm ICEC}(\varepsilon)$ for several values 
of {\it R}. We see that below the ICEC threshold only PR is possible, whose 
cross section is about 10$^{-4}$ Mb. Above threshold, the ICEC channel opens 
and its cross section at the rather large interatomic distance of 1 nm 
is more than three orders of magnitude larger than that of PR for the 
energies in question. Although the ICEC cross section falls off fast 
with {\it R}, ICEC remains the dominant process even at {\it R} as large as 3 nm.

We consider next electron capture by Cl having Br$^{-}$ as a neighbor. Here, 
obviously, the ICEC threshold is $\varepsilon^t$=0 and interatomic electron capture 
can take place at all energies of the free electron. The calculated PR 
and ICEC cross sections shown in Fig.\ref{fg1}C demonstrate that ICEC again 
dominates over PR up to interatomic distances as large as 3 nm. The ICEC 
cross section is largest at threshold and at a distance of 1 nm between 
Cl and Br$^-$ it takes close to the threshold the respectable value of $\approx$0.4 Mb.

\begin{figure}[htb]
\begin{center}
\includegraphics[width=12.cm]{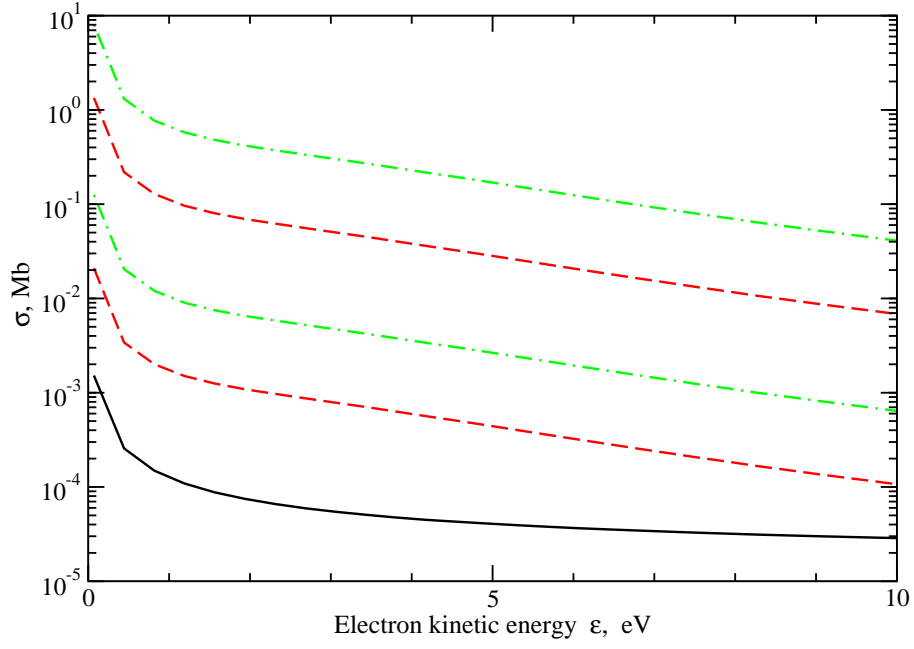}
\caption{PR cross section of Mg$^{2+}$ and ICEC cross sections of Mg$^{2+}$ with N water molecules as neighbors:
solid line - $\sigma_{\rm PR}^{{\rm Mg}^{2+}}(\varepsilon)$; dashed lines - $\sigma_{\rm ICEC}$ of
Mg$^{2+}$ with one H$_2$O at $R_{\rm Mg-O}$=5 \AA (upper curve) and at $R_{\rm Mg-O}$=1 nm (lower curve); 
dashed-dotted lines - $\sigma_{\rm ICEC}$ of Mg$^{2+}$ with 6 H$_2$O at 
$R_{\rm Mg-O}$=5 \AA (upper curve) and at $R_{\rm Mg-O}$=1 nm (lower curve).}
\label{fg2}
\end{center}
\end{figure}

The above examples were concerned with the formation of negative ions by 
electron capture. Of course, it is of great interest to consider 
electron capture by positive ions as well. Here, typically, the energy of 
the virtual photon transferred in the process is much larger, and, 
since $\sigma_{\rm ICEC}\sim E_{vph}^{-4}$, the question arises immediately, whether in 
environment ICEC can still be more important than PR. Alkali and 
rare-earth metal cations in the presence of a water molecule 
constitute examples of general interest and are at the same time 
critical choices, since the energy of the virtual photon must be larger 
than the rather large IP of water (the IP of an isolated water molecule 
is 12.62 eV). To be specific we take the metal cation to be Mg$^{2+}$ and 
since EA$_{{\rm Mg}^{2+}}$=14.74 eV \cite{ivanov1994}, the minimal value of $E_{vph}$ is also 14.74 eV, which is about 
five times larger than that in the above examples. As usual, the
PR cross section of Mg$^{2+}$ is obtained from the photoionization cross 
section of Mg$^+$ \cite{ivanov1994}, using $g_{{\rm Mg}^+}$=2 and $g_{{\rm Mg}^{2+}}$=1. The photoionization cross 
section of water needed to compute $\sigma_{\rm ICEC}$ is taken from \cite{haddad1986}. The results 
for $\sigma_{\rm ICEC}$ are collected in Fig.\ref{fg2}, where $\sigma_{\rm PR}^{({\rm Mg}^{2+})}$ is also shown for 
comparison. We see that at {\it R}=5 \AA, which is more than 2.5 times larger 
than the equilibrium Mg-O distance of 1.94 \AA~ in Mg(H$_2$O)$^{2+}$ \cite{pavlov1998}, 
the ICEC cross section is 10$^2$-10$^3$ times larger than the PR cross 
section. Close to threshold $\sigma_{\rm ICEC}$ reaches values which are above 1 Mb. 
At a distance of 1nm the ICEC cross section is still clearly larger than the PR cross section (see Fig.\ref{fg2}). 
Since one can easily choose ions and neighbors which are more favorable for ICEC than Mg$^{2+}$ and water, 
we may conclude that in spite of the large virtual photon energies ICEC can dominate over PR even at ion-neighbor 
distances of a few nanometers.

So far we considered examples where the electron is captured in the 
presence of a single neighbor B. What happens if several neighbors B$_i$ 
are available? For instance, in a chemical or biological
environment a Mg$^{2+}$ ion will usually be in the presence of several, if not 
many, water molecules. The excess energy can now be transferred to any 
one of the neighbors and ionize it. If these neighbors are 
sufficiently far from each other, one can neglect the interaction among 
the various virtual photon transfer channels. The total ICEC cross 
section is then nothing but the sum of the individual ICEC cross 
sections of A in the presence of a neighbor B$_i$ and reads
\begin{equation}
\sigma_{\rm ICEC}(\epsilon)=\sum_{i}P_{i}(E_{vph},R_{i})\sigma_{\rm PR}^{\rm (A)}(\epsilon),
\label{eq5}
\end{equation}
where the summation runs over all neighboring species contributing to 
ICEC. In the case of N equidistant neighbors of the same type this 
expression becomes $\sigma_{\rm ICEC}(\epsilon)=NP(E_{vph},R)\sigma_{\rm PR}^{\rm (A)}(\epsilon)$.
Since N can easily be of the order of 10, the presence of a surrounding 
medium might lead to an increase in $\sigma_{\rm ICEC}$ by an order of magnitude. 
To demonstrate the enhancement of ICEC due to several neighbors, we plot in Fig.\ref{fg2} $\sigma_{\rm ICEC}$ 
of Mg dication surrounded by 6 water molecules which are located at R=5\AA~and at R=1nm from Mg$^{2+}$. 
These cases have been merely chosen as illustrative examples to demonstrate the enhancement of ICEC 
due to the presence of several neighbors. As mentioned above, one can easily choose ions and neighbors 
more favorable for ICEC at even larger ion-neighbor distances.

From the above analysis we conclude that in the presence of 
neighboring species PR is accompanied by a new interatomic process 
where the electron capture proceeds in concert with the ionization of a 
neighboring species. The explicit expression of the ICEC cross section 
derived in the framework of the virtual photon transfer model and valid 
for large interatomic separations allows one to determine the conditions 
under which ICEC dominates over PR. ICEC will be preferable for 
slow electrons and/or in systems with low electron affinities and 
ionization potentials of the participating species. Examples of the 
latter are provided by halogen/halide systems, where ICEC dominates 
over PR up to interatomic distances of several nanometers. The 
interatomic process is also preferred in systems where the
ionizable moiety possesses large photoionization cross sections close to 
the photoionization threshold, e.g. in the form of shape resonances. 
This makes molecules more attractive neighbors than atoms. But perhaps 
the most important conclusion is that in a medium the capturing center 
will usually have more than one neighbor with whom it can undergo ICEC 
efficiently. To a first approximation one can simply add up the 
individual cross sections of the capturing center with each of the 
neighbors, thus strongly enhancing the resulting total ICEC cross section.

The interatomic process we have discussed above is a new fundamental process with its own merits. 
We compared its cross section with that of PR, which is the electron capture mechanism for isolated atoms, 
in order to have a quantitative reference for the impact of a possible environment.
If one wants ICEC to dominate, the systems of interest should have densities which correspond to interatomic 
distances between nearest neighbors of up to a few nanometers. 
The desired conditions are amenable to experiments in the laboratory.
By varying the density and the neighboring species, the impact of the environment on electron capture can be 
systematically explored. Moreover, by investigating ICEC one can study interatomic correlation effects which are 
of interest by themselves. It is expected that ICEC plays a role also in nature, but being a newly discovered 
mechanism, details of this role must be left to future studies. For reference we mention that the average distance 
between air molecules at sea level is about 2 nm. In the stratosphere and in commonly encountered plasmas the 
density is much lower making PR the dominant capture mechanism.
Last but not least, we also mention biological systems where positive and negative chemically unbound ions are in 
favorable distances from neutral species and an abundance of slow electrons 
is provided by high energy radiation \cite{mozejko2003}.

We hope that the present results will stimulate experiments and further 
theoretical developments.

\bibliographystyle{prsty}

\newpage

\end{document}